\documentclass[prl,twocolumn,showpacs,preprintnumbers]{revtex4}
\usepackage{amsmath,amssymb,latexsym}

\usepackage{graphicx}

\begin{document}

\title
{The dynamic conductivity  and the plasmon profile of Aluminum
in the ultra-fast-matter regime.
}

\author
{
 M.W.C. Dharma-wardana}
\affiliation{
National Research Council of Canada, Ottawa, Canada, K1A 0R6
}
\email[Email address:\ ]{chandre.dharma-wardana@nrc-cnrc.gc.ca}

%
\date{\today}

\begin{abstract} 
 We use an explicitly isochoric  two-temperature theory to analyze recent
X-ray laser scattering data for Aluminum in the ultra-fast-matter (UFM)
regime up to 6  eV. The observed
  surprisingly low conductivities are explained by including strong 
electron-ion scattering  effects  using the phase shifts calculated via
 the neutral-pseudo-atom model. 
 The applicability of the Mermin model to
UFM is questioned.
 The static and dynamic conductivity,  complex collision frequency
 and the plasmon line-shape are evaluated within a  Born approximation and are
in  good agreement with experiment. 
\end{abstract}

%
\maketitle
\section{Introduction}
{\it Introduction -}
\label{intro} 
Short-pulsed X-ray photons, e.g., from the Linac Coherent Light Source
(LCLS) have begun to  provide data in hitherto inaccessible regimes of
matter~\cite{Sper2015, GlenzerRMV}. Such information is of interest in
 understanding
normal matter under extreme
conditions~\cite{Ng2011,chen2013,PNAS,Milchberg88}, as well as at
 new frontiers in high-energy-density matter, astrophysics,
fusion physics etc. Such non-equilibrium systems are also produced in
semiconductor devices~\cite{DharmaHot93}. The
theory involves complicated many-body effects and the
quantum mechanics of finite-temperature non-equilibrium systems.
Standard {\it ab-initio} methods are inapplicable or
computationally prohibitive for this ultra-fast matter (UFM) regime. 
Extensions of elementary plasma models or Thomas-Fermi
models fail badly. Hence computationally simple realistic theories
  of these systems are essential in
 the interpretation of experiments on UFM which
 is a sub-class of  warm-dense-matter (WDM)~\cite{cimarron}. Here we use
 a finite-$T$ density-functional theory (DFT)
 calculation of the electronic charge distribution $n(r)$
 and the ion charge distribution $\rho(r)$ around an Al ion in the system as
the basic ingredient of such a theory. The
neutral pseudoatom (NPA) model of  Perrot and 
Dharma-wardana~\cite{eos95,CPP} is used in this study.

The LCLS results~\cite{Sper2015} of the plasmon feature and the dynamic and
static conductivities $\sigma$ of Al up to 6 eV, isochorically held at  solid
density dramatically improves  on the accuracy of the earlier UFM experiments
~\cite{Milchberg88,pwd-milsch}. Surprisingly low static conductivities $\sigma(0)$
of  UFM aluminum are reported in Ref.~\cite{Sper2015}, even at 0.2
eV.

We present two-temperature (2$T$) calculations for isochoric
Aluminum.  Atomic units (a.u., $|e|=\hbar=m_e=1$) are used, and the
temperature is in energy units. The ion temperature $T_i$ is the
initial `room' temperature, while only the electron temperature $T_e$
 is raised to 6 eV by the
50 femto-second X-ray  pulse. We do not get the gradual decrease of
$\sigma$ with  $T$ found for  equilibrium
 non-isochoric aluminum. Instead, we reproduce
the low static conductivities reported in the experiment. The  high
conductivities of the normal solid and the  molten metal ($T_e=T_i$) at
low $T$  are partly attributed to 
the position of the scattering momentum $2k_F$  {\it falling within
the second minimum} in the ion-ion structure factor $S(q)$. In an
isochoric UFM solid, the ions have no time to adjust to the rapidly
heated electrons. The ions  (and their bound electrons) remain frozen
at their lattice sites, and at $T_i$. Hence $S(q)$, and the bare
electron-ion pseudopotential $W(q)$ remain essentially unchanged,
even up to $T_e=6$ eV. The thermal smearing of the Fermi sphere is set
by $f'(k,T_e)=f(k,T_e)(1-f(k,T_e))$, where $f(k,T_e)$ is the electron Fermi
function. It's  overlap with the ion-ion  $S(q)$, and  the electron-ion
scattering cross section   determine
the conductivity  $\sigma(0)$ as well as $\sigma(\omega)$.

The new experiment provides the profile of the plasmon
resonance.  We present a simple theory of the
momentum relaxation and energy dephasing frequency $\nu(\omega)$
(also known as the `collision frequency'),  using a
Born approximation constructed to match the $\omega\to 0$ 
conductivity obtained from the NPA phase shifts. The calculated plasmon
profile is in good accord with experiment.

{\it The DFT-NPA model for isochoric UFM Aluminum -} An aluminum
nucleus is placed in an electron subsystem and an ion
subsystem, within a  large sphere  ($R\sim$ 30 au.) where  all particle
correlations  reach bulk values as $r\to R$.  
Hence this `neutral-pseudo-atom' (NPA) is not an  `average-atom cell-model'
 similar to the INFERNO model of Lieberman or its 
 improvements~\cite{Lieberman}.
The electron density in the bulk, viz., $n_e$ is 1.81$\times 10^{23}$
electrons/cm$^3$, has an electron-sphere radius
$r_s=2.07$ au. The  free-electron pile up $n_f(r)$  and the
scattering phase shifts $\delta_{kl}$ around the Al nucleus  are
calculated via the Kohn-Sham equations, using a step-function
 to  mimic the ion-ion pair distribution
 function  $g(r)$.
 This is known to work well for Al~\cite{eos95}. 
The phase-shifts satisfy the Friedel sum rule, and the DFT 
uses a finite-$T$ exchange-correlation
contribution~\cite{pdw-exc}. All the results in this study
follow from the NPA output. The many-ion
system is built up via the  $S(q)$ as a
 superposition of NPAs, using the $S(q)$ derived within the
theory. 

\begin{figure}
\includegraphics*[width=8.5 cm, height=6.0 cm]{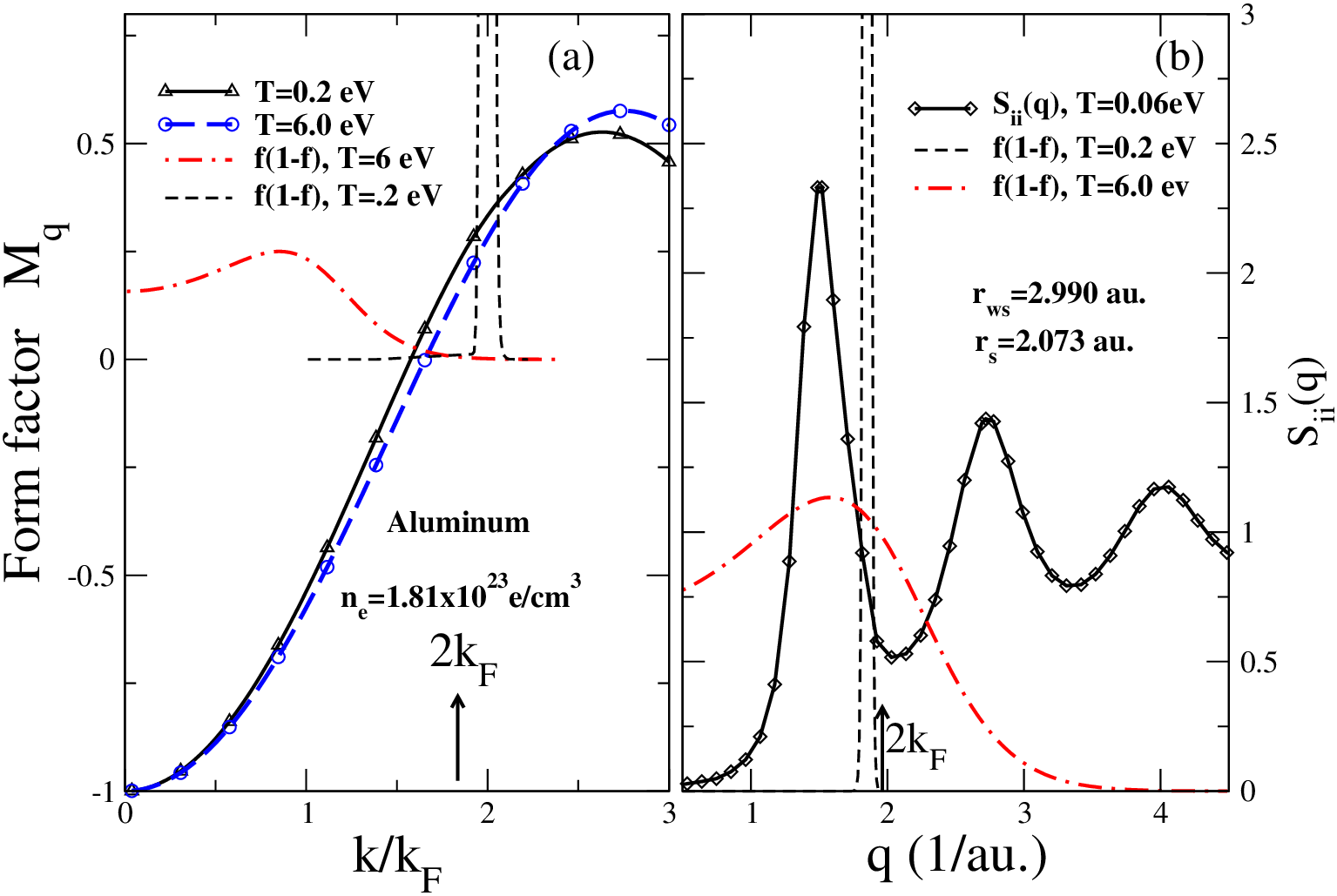}
 \caption
{(Online color). (a) The pseudopotential form factor $M(q)$ at
 $T=0.2$ eV and 6 eV, and the thermal-smearing
 functions $f'(k,T_e)=f(k)(1-f(k))$. (b)
 The overlap of $S(q)$ and $f'(k,T_e)$. The ion $S(q,T_i)$ with $T_i$ = 0.06 eV.} 
\label{nq-Fig}
\end{figure}

At room temperature, this calculation yields an ionization $Z=3$ and
an ion Wigner-Seitz radius $r_{ws} \simeq $ 2.99 au. The $r_{ws}$ is
 held constant while $T_e$ is increased, to mimic the isochoric
UFM, where as normal solid or liquid Al
expands (i.e, $r_{ws}$ increases) with temperature.  A static electron
response function $\chi(q,T_e)$ is constructed, with its local field
correction (LFC)  satisfying the compressibility sumrule at
each temperature. This defines a fully local pseudopotential
$W(q)=n_f(q)/\chi(q,T_e)$, and an ion-ion pair potential
$U_{ii}(q)=Z^2V_q-|W(q)|^2\chi(q)$. The
pseudopotential $W(q)=-ZV_qM_q,\; V_q=4\pi/q^2$ is fitted to a
Heine-Abarenkov form for convenience. The form
factor $M_q=n_f(q)/n^0_f(q)$ obtained from the NPA is shown in
 Fig.~\ref{nq-Fig}(a)  at $T=$0.2 and 6 eV.  Here $n_f^0(q)$ is the
 linear-response charge pileup.  
This approach is capable of milli-volt accuracy and 
reproduces even the high-temperature phonons~\cite{Harbour-CPP}
 discussed by, e.g.,  Recoules et al~\cite{Recoules2013} 
(but phonons do not  form during  UFM timescales).

The resulting $U_{ii}(q)$ is used in the modified Hyper-Netted-Chain
equation (MHNC) yielding the  $S(q)$ at the
ion temperature $T_i$ (which is the initial temperature of the system
at the arrival of the X-ray pulse). Since the
initial  Al-crystal has an FCC structure, it is sufficient to use the
spherically averaged $S(q)$  taken as
a `frozen fluid', say, at 0.06 eV.  The latter is the lowest temperature at which
the HNC could be converged, since the melting point is $\simeq$ 0.082 eV. The
results are insensitive to the use of an $S(q)$ at 0.06 eV or, say, 0.082
eV. Our MHNC procedure is accurate enough to
closely reproduce the experimental $S(q)$ of
normal liquid aluminum~\cite{cdw-aers-Al}. 

{\it The complex conductivity $\sigma(\omega)$ - }
 The Drude theory with a static $\nu(0)$ is  known to be
 inadequate for $\sigma(\omega)$ except at
 small and high $\omega$~\cite{Hopfield}.
Sperling et al~\cite{Sper2015} have used a Mermin model (diffusion pole)
~\cite{Mermin}  augmented by plasma
many-body theory~\cite{Reinholz2000} where they combine components of Born (B),
Lenard-Balescu (LB)  and Gould-DeWitt (GDW)-Mermin (M) approaches in their analysis
where $T_i=T_e$.
The real part of the complex conductivity
 $\sigma(\omega)=\sigma_1+i\sigma_2$, obtained
via B-LB-GDW-M  is two orders of magnitude too large
 compared to experiment,  although the imaginary part $\sigma_2(\omega)$
 as well as the plasmon profile are in much better accord. They
use several models of $S(q)$, point-ion Coulomb potentials as well
 as pseudopotentials. Since
the $\omega\to 0$ limit of the $\sigma(\omega)$  gives 
a poor  $\sigma(0)$, they use a Ziman formula with suitable
models of  $S(q)$ and pseudopotentials.

In our approach, the ion-$S(q,T_i)$ at $T_i$  remains intact for
all $T_e$. We first calculate $\sigma(0)$ using the electron phase shifts obtained
from the NPA  and obtain good agreement with experiment. The calculation
of $\sigma(\omega)$ via the phase shifts is more demanding.
Instead, since Al is a ``simple metal'',
an Ashcroft pseudopotential 
$V_A(r_c)$  specified only by the core radius $r_c$ that reproduces the $\sigma(0)$ 
could be found. This $r_c$ is
consistent with the NPA value. This is  used in calculating $\sigma(\omega)$.
There is no low-frequency `diffusion pole' in the experimental
spectra as expected from  Mermin theory. Mermin assumes that the ions respond
perfectly to the electron-density fluctuations and maintain local charge
neutrality. This holds for  timescales $t$
much larger than the electron-ion temperature relaxation time $\tau_{ei}$ which is
many pico-seconds~\cite{elr2001} if $T_i\ne T_e$, or for timescales
significantly larger than phonon timescales if $T_i=T_e$. Thus the Mermin model
is largely inappropriate for most UFM-WDM systems. 
Hence we examine  a simple  RPA-like model where the ions are mere
immobile scatterers during the 50 fs signal,  and obtain
good  overall agreement with experiment.

The conductivity $\sigma(\omega)$ can be expressed via the force-force
 correlation function as given in standard texts~(e.g.,
 Ref.~\cite{Vignale2005} sec.~4.6). If plane waves
are used for  the free electrons, 
the limit $\omega\to 0$ recovers  the Ziman formula. However, if the electron-ion
interactions are strong, then
the electron response  $\chi(q,\omega)$ and the dynamic conductivity 
$\sigma(\omega)$  should be  expressed  via the electron
 eigenstates $\phi(r)_{\alpha}$ of the system~\cite{Ehren63,Dharma78}.
The NPA provides these, with $\alpha=n,l$ for core-states,
 and $\alpha=k,l; E_{kl}=k^2/2$
for continuum states, with the $m$ quantum number and spin
summed over~\cite{Ehren63,Dharma78}. The core states give bound-bound
transitions, while the bound-continuum and continuum-continuum transitions are
also included. If numerical eigenstates $\phi_\alpha(r)$ are
not available, hydrogenic functions can be used within a
many-body theory as  in Ref.~\cite{Dharma78}. Such
 ``Green-Kubo''  formulae for $\sigma(\omega)$ usually need
 heavy numerical  codes. Our NPA approach gives a  simpler evaluation
 of comparable accuracy with orders of magnitude rapidity.
\begin{figure}
\includegraphics*[width=8.5 cm, height=11.0 cm]{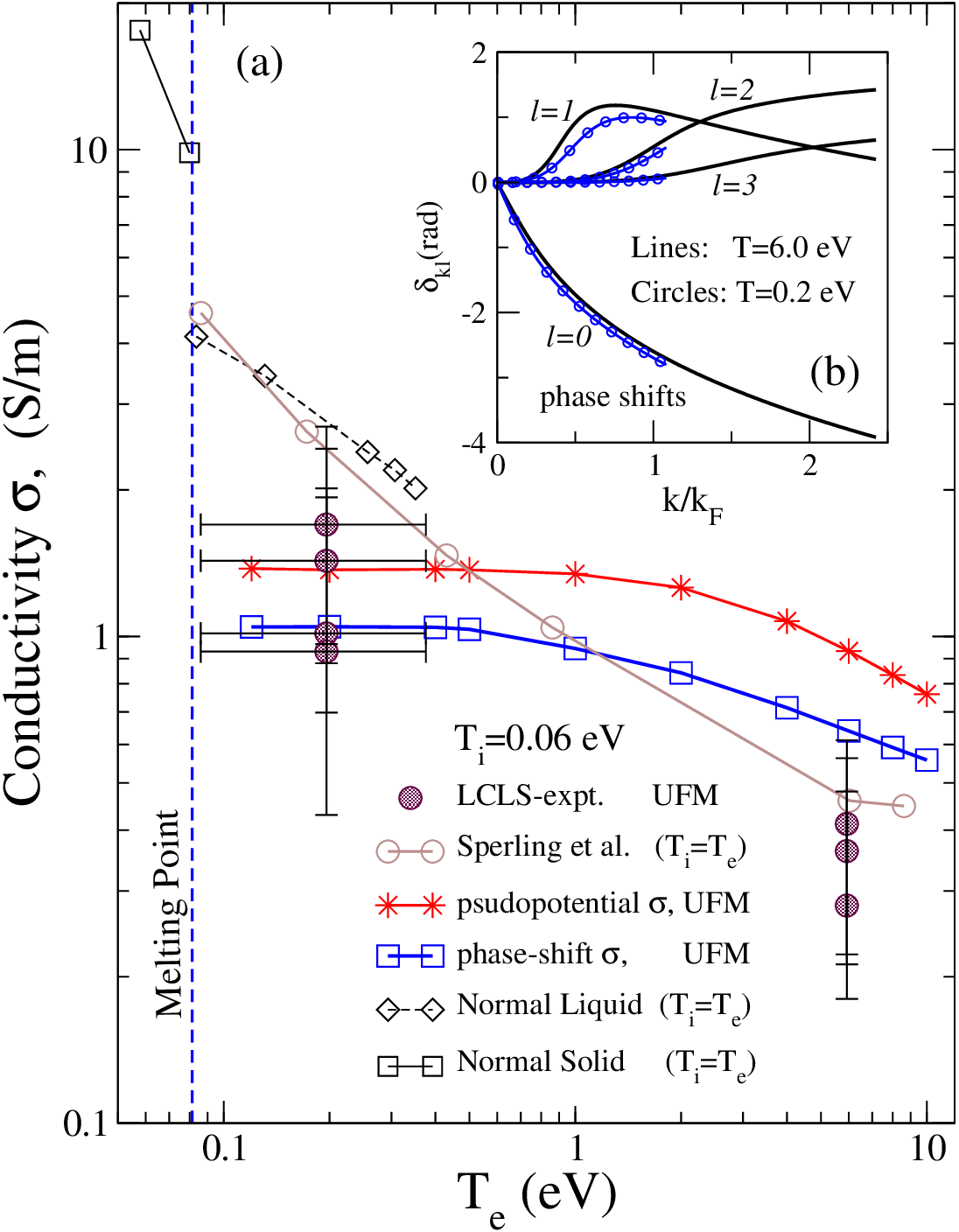}
 \caption{(Online color) (a) The static conductivity $\sigma(0)$ of
Aluminum. LCLS experiment and the $\sigma(0)$ from theory for
UFM aluminum ($T_i\ne T_e$). Some data for the normal
solid and normal liquid are also shown. Sperling et al($T_i=T_e$) data are
a private communication.
(b) The NPA phase shifts $\delta_{kl}$ are shown for
$l=0$-3, as a function of $k/k_F$.}
\label{cond-fig}
\end{figure}
The corrections beyond the non-interacting response can be
expressed as a relaxation frequency $\nu(\omega)=\nu_1+i\nu_2$
given in terms of a scattering cross section. The $\nu$ describes
momentum relaxation as well as energy dephasing. This can be
expressed via the NPA phase shifts~\cite{pdw-res,Thermophys}.
The real part $\nu_1(\omega)$ may be given as:
\begin{eqnarray}
\label{dyn-resp.eqn}
\nu_1(\omega)&=&\frac{\Im}{3Z}\sum_{\vec{q},\vec{k}}
q^2\Sigma(\vec{k},\vec{q})S(q)\frac{f(\vec{k})-f(\vec{k}+\vec{q})}
{i\omega(\omega+\epsilon_{\vec{k}}-\epsilon_{\vec{k}+\vec{q}})}\\
\label{cross-sec.eqn}
\Sigma(k,q)&=&\left|k^-2\sum_l (2l+1)e^{i\delta_{kl}}
\sin(\delta_{kl})P_l(cos\theta)\right|^2
\end{eqnarray}
The static limit of Eq.~\ref{dyn-resp.eqn} gives:
\begin{eqnarray}
\label{sig0-eqn}
\nu(0)&=&\frac{1}
{ 3\pi Z T_e}\int_0^\infty f(k)(1-f(k) k^2 dk F(k) \\
F(k)&=&\int_0^{2k}q^3\Sigma(q,k)S(q)dq;\;q=k(1-cos\theta)^{1/2}
\end{eqnarray}
The original numerical implementation (see appendix, Ref.~\cite{pdw-res})
has been improved, using up to 38 $l$-states if needed,
using an energy cutoff of $E_F+2T_e$, together  with
asymptotic corrections.  Typical  $\delta_{kl}$  from the
NPA are shown in fig.~\ref{cond-fig}(b).
Results for  $\sigma(0)$ for
 isochoric Al, from Eq.~\ref{sig0-eqn}
covering 0.2 eV to 10 eV  are given in
Fig.~\ref{cond-fig}(a) while $\sigma(0)$  up to 100 eV are in Table.1 of
Ref.~\cite{Thermophys}. If the scattering cross section is evaluated
using plane waves (i.e., Born approximation), Eq.~\ref{sig0-eqn}
reduces to the Ziman formula  with the weak
 pseudo-potential $W(q)=ZV_qM_q$ (shown in Fig.~\ref{nq-Fig}).
The Heine-Abarenkov $W(q)$ gives a higher estimate of  $\sigma(0)$, 
while the phase-shift calculation agrees with LCLS.

We replace  $W(q)$ by an  Ashcroft pseudopotential 
 $V_A(q)$  chosen to {\it reproduce the static conductivity} $\sigma(0)$,
 and use it  to evaluate the 
relaxation frequency $\nu(\omega)$
in the Born approximation to Eq.~(\ref{dyn-resp.eqn}). Thus,
\begin{eqnarray}
\label{sig-dyn.eqn}
\nu(\omega)&=&\frac{1}{6\pi^2Z}\int q^4|V_A(q)|^2S(q,T_i)
\Delta(q,\omega) dq\\
\label{sig-dyn2.eqn}
\Delta(q,\omega)&=&\frac{\{\chi_e(q,\omega,T_e)-\chi_e(q,0,T_e)\}}{i\omega}
\end{eqnarray}
Eq.~(\ref{sig-dyn.eqn}) is basically  Hopfield's expression~\cite{Hopfield}, 
while modern discussions are found in Refs.~\cite{Mahantxt,Reinholz2000}. 
 The $S(q)$ is for the
cold ions at $T_i=0.06$ eV, as shown in Fig.~\ref{nq-Fig}(b).
\begin{figure}
\includegraphics*[width=8.5 cm, height=6.6 cm]{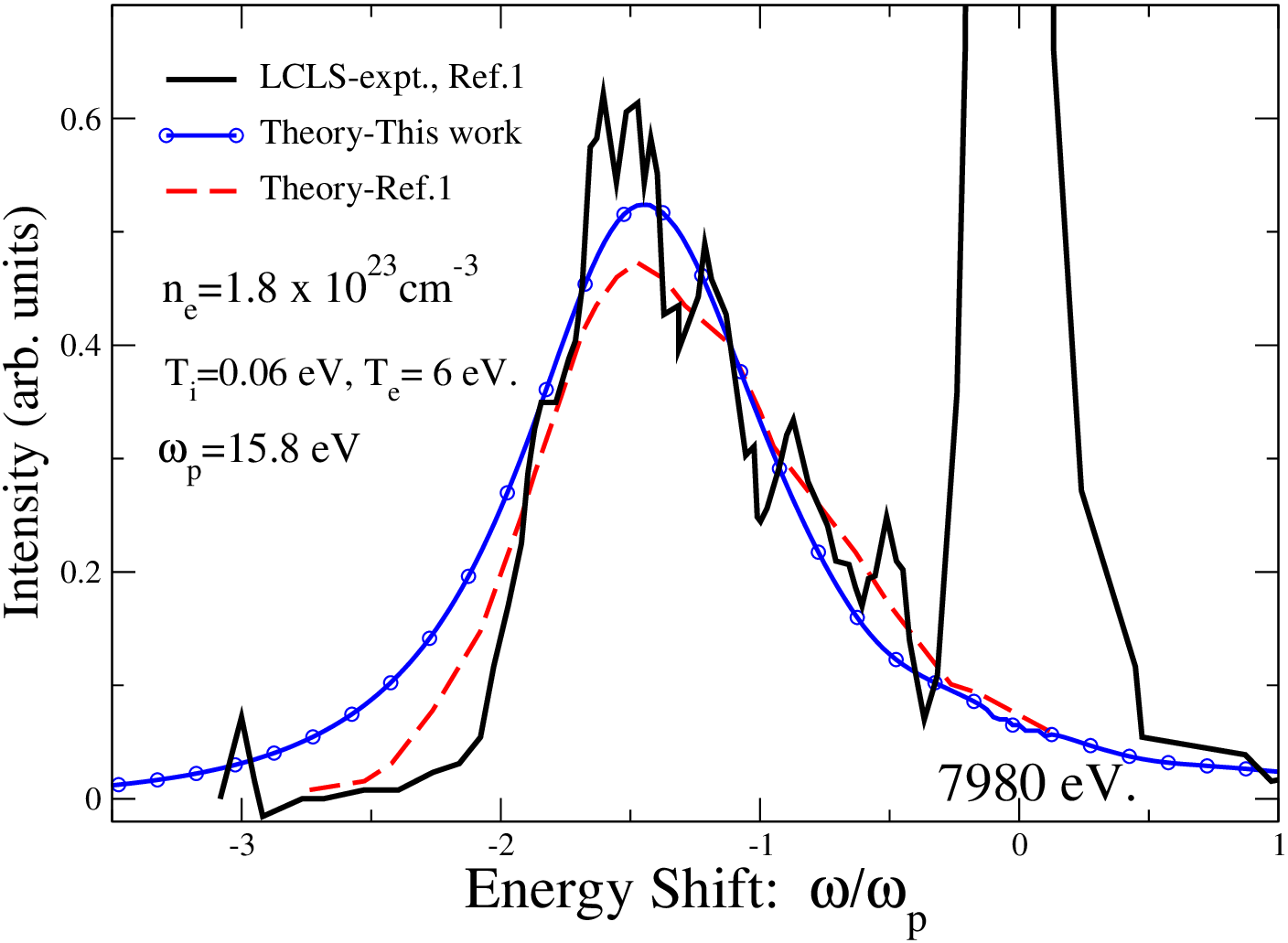}
 \caption{(Online color) The UFM-aluminum plasmon lineshape
at $T_e= 6$ eV,  from experiment and theory.}
\label{plsmon.fig}
\end{figure}

{\it The plasmon profile and $\nu(\omega)$.} 
\label{dyn-cond} 
An important result of the LCLS-experiment is the plasmon profile from
UFM-Aluminum. We discuss  $T=6$ eV in detail. Eq.~(\ref{sig-dyn.eqn})
evaluates  $\nu_1(\omega)$ and $\nu_2(\omega)$ using the
$V_A(r_c)$  pseudopotential. Obtaining $\nu_1$ via $\Im\{\chi(q,\omega)\}$ in
 Eq.~\ref{sig-dyn2.eqn} and $\nu_2$ via  Kramers-Kronig is
 computationally convenient. A direct
estimate of $\nu_2$ is also available from 
Eqs.~(\ref{sig-dyn.eqn}) and (\ref{sig-dyn2.eqn}).
The  response function $\chi(q,\omega)$
uses an LFC  derived from the finite-$T$ xc-potential~\cite{pdw-exc}. The
transverse dielectric function $\varepsilon(q\to 0, \omega+i\nu(\omega))$
provides  the optical scattering cross section $S_{ee}(q\to 0,\omega)$.
 This is
$\propto \Im\{1/\varepsilon{(\omega-\nu_2+i\nu_1)}\}n_B(\omega)$ where
 $n_B(\omega)$ is a Bose factor at the electron temperature $T_e$.
Instead of Mermin theory we use the simple RPA-like transverse dielectric
 function in the $q\to0$  limit.
 The  calculated scattered intensity 
is shown in Fig.~\ref{plsmon.fig}. The  predicted profile differs 
on the red wing of the experimental plasmon line shape. Since $V_A(r_c)$ was
 fitted to the phase-shift $\sigma$ only at $\omega=0$, this  is not surprising.

\begin{figure}
\includegraphics*[width=8.5 cm, height=6.6 cm]{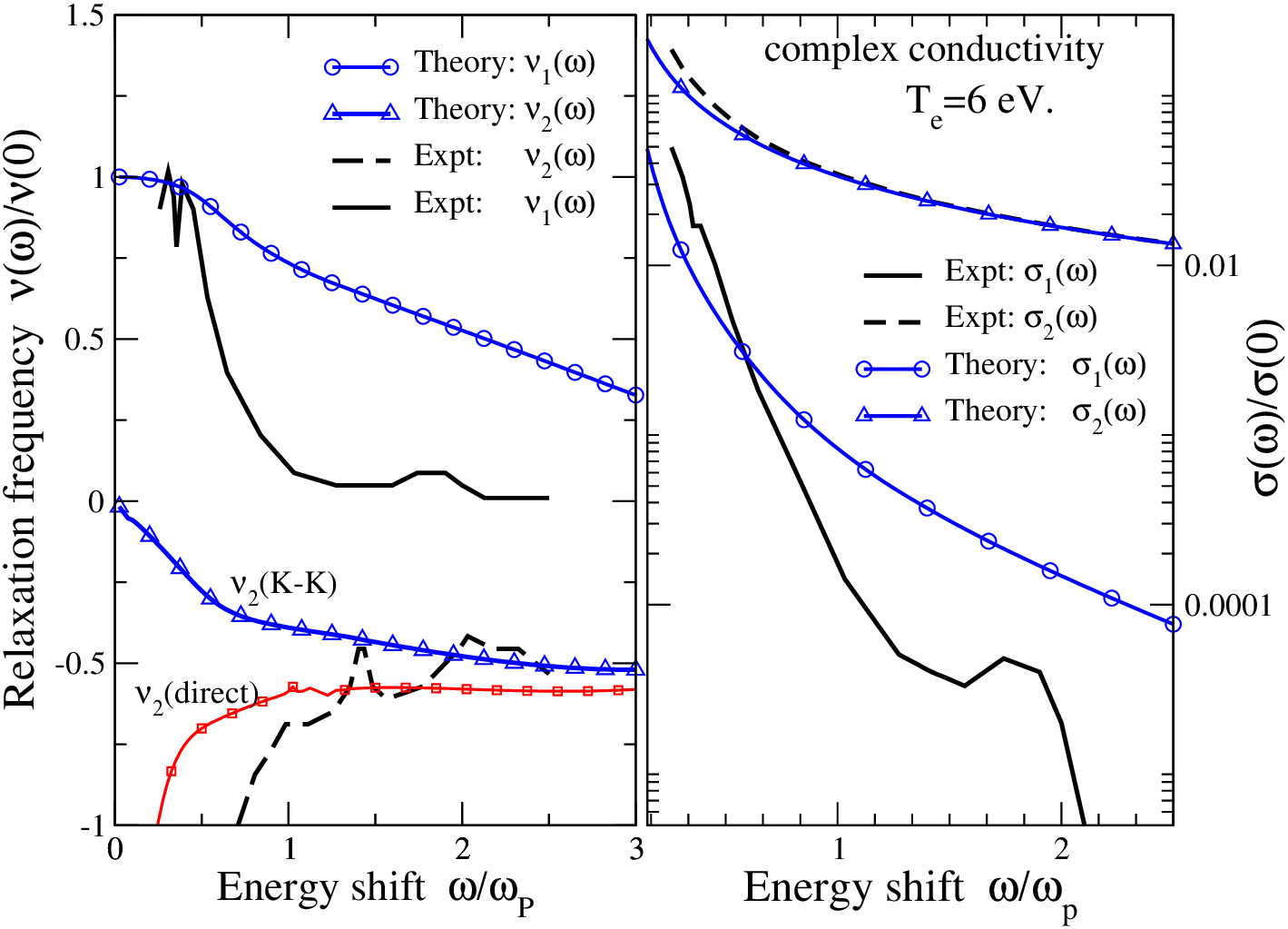}
 \caption{(Online color) (a) Theory and experiment for the  momentum-relaxation
 frequency $\nu_1$ and $\nu_2$ versus the energy shift $\omega/\omega_p$.
$\nu_2(\omega)$ calculated from $\nu_1(\omega)$ via Kramers-Kr\"{o}nig, and via a
direct numerical procedure are shown.
 (b)
$\sigma_1(\omega)$ and $|\sigma_2(\omega)| $ from experiment and theory.}
\label{sig-om.fig}
\end{figure}

The relaxation frequency $\nu(\omega)$ and the conductivity $\sigma(\omega)$
can be extracted from  the experimental  $S(q\to 0,\omega)$. 
We use the experimental $\nu_1(\omega),\nu_2(\omega)$ of Sperling
et al., to test our methods, even though they assumed a  Mermin form
to extract the data, assuming that the modeling differences fall within the error bars.
 The experimental $\nu^{ex}_1$ and $\nu^{ex}_2$ are compared  with the
 calculated $\nu_1,\nu_2$
in the figure~\ref{sig-om.fig}, where the energy shift $\omega$ is 
$\omega_1-\omega_0$ with $\omega_0$=7980 eV., and hence negative (for the plasmon
studied here). The theoretical $\nu_1$ decays very slowly compared the $\nu^{ex}_1$.
We expect this to be corrected when a full evaluation using phase shifts is used.
 
The  Drude formula provides $\sigma(\omega)$ from $\nu(\omega)$. 
Setting $\varpi=\omega-\nu_2,\,
d(\omega)=\nu_1(\omega)^2+\varpi^2$ we use $\alpha=\omega_p^2/(4\pi),\,
 \sigma_1(\omega)=\alpha\nu_1/d$ and $\sigma_2(\omega)=\alpha \varpi/d$. We have
recalculated $\sigma_1,\sigma_2$ from our theoretical $\nu_1,\nu_2$ given in
Fig.~\ref{sig-om.fig}(a), and from the experimental $\nu_1,\nu_2$ at $T$=6 eV given in
Fig.~3 of the supplementary material of Ref.~\cite{Sper2015}. 
The resulting $\sigma_1(\omega),\sigma_2(\omega)$
are displayed in Fig.~\ref{sig-om.fig}(b). Note that although $\sigma_2(\omega)$ is expected to
tend to zero as $\omega\to 0$, this happens only quite close to $\omega=0$ because of the
strong negativity seen in both experimental and theoretical numbers for $\nu_2$
(see Fig.~\ref{sig-om.fig}(a)). 
Although $\sigma_1$ is close to the experiment for small-$\omega$, it begins to
differ significantly from experiment as $\omega$ increases.
 
In conclusion, the  static conductivity calculated using
phase-shifted NPA electron eigenfunctions  
for  two-temperature ultra-fast aluminum are in good
agreement with the LCLS data. A simple Born approximation
to the  dynamic conductivity using a pseudopotential fitted to the
theoretical $\sigma(0)$ provides a good approximation to the
plasmon lineshape and the dynamic conductivity obtained from the
LCLS experiment. It is argued that the Mermin form is inappropriate for
ultrafast matter where the ions have no time to respond. 
A full calculation of $\nu(\omega)$ entirely from the
phase shifts via Eq.~(\ref{dyn-resp.eqn}) may resolve some of the
 shortcomings in the present theory. 

The author thanks Heide Reinholz, Philipp Sperling and  colleagues for
their  comments.

\end{document}